\documentstyle[prl,aps,epsf,multicol]{revtex}
\begin{document}
\draft
\tighten

\title{Nematic--to--Smectic-A Transition in Aerogel} 
\author{Leo Radzihovsky} 
\address{Department of Physics, University of Colorado,
Boulder, CO 80309} 
\author{John Toner} 
\address{Dept. of Physics,
Materials Science Inst., and Inst. of Theoretical Science,
University of Oregon, Eugene, OR 97403}

\date{\today}
\maketitle
\begin{abstract}
We study a model for the Nematic--Smectic-A (NA) transition in
aerogel, and find that even arbitrarily weak quenched disorder (i.e.,
low aerogel density) destroys translational (smectic) order.  Ignoring
{\em elastic} anharmonicities, but keeping anharmonic couplings to
disorder, leads to the prediction that there is no ``Bragg glass''
phase in this system: it is riddled with dislocation loops induced by
the quenched disorder.  Orientational (nematic) order is destroyed as
well, as is the thermodynamically sharp NA transition, in agreement
with recent experimental results.
\end{abstract}
\pacs{64.60Fr,05.40,82.65Dp}

\begin{multicols}{2}
\narrowtext

Recent experiments\cite{Clark,HG} have focussed on the study of liquid
crystals in the random environment of an aerogel, as new paradigm
system in which to investigate the general problem of quenched
disorder in condensed matter systems.

In this Letter we develop a theory of such systems.  We find that in a
harmonic {\em elastic} approximation, even arbitrarily weak disorder
destroys both the smectic-A phase and the NA transition, and creates
unbound dislocation loops.  The formalism we use to demonstrate this
defect unbinding is new, powerful, and potentially applicable to a
wide variety of candidate ``Bragg glass''\cite{BraggGlass} systems.

Once dislocations are present, the phase is best characterized as a
nematic in a random tilt field.  However, subsequent examination of
orientational fluctuations in this nematic leads to the conclusion
that tilt disorder destroys the orientational order of the smectic
layers as well.
 
However, a ``nematic Bragg glass'' and a thermodynamically sharp
"nematic glass" transition may occur\cite{toPublish}, in agreement with
recent dynamic light scattering experiments\cite{Clark} that show a
dramatic slowing down of director fluctuation relaxations in liquid
crystals in aerogel below a temperature $T_g$ near the bulk NI
transition.  

Near the NA transition the center-of-mass nematogen molecular density
$\rho({\bf r})$ develops strong fluctuations dominated by Fourier
components near the smectic ordering wavevector ${\bf q_0}={\bf n}
2\pi/d $ parallel to the nematic director ${\bf n}$.  Defining the
local (complex scalar) order parameter $\psi({\bf r})$ which
distinguishes the smectic-A from the nematic phase\cite{deGennes}
via:$\,\rho({\bf r})= \mbox{Re}[\rho_0+e^{i {\bf q_0}\cdot{\bf
r}}\psi({\bf r})]$, where $\rho_0$ is the mean density of the smectic,
we take for our free energy:$\,F=F_{dG}+F_{d \rho}+F_{dn}$, where
$F_{dG}[\psi,{\bf n}]$ is de Gennes free energy (which includes the
Frank free energy)\cite{deGennes}, and $F_{d \rho}$ and $F_{d n}$ are
the new disorder parts, which couple the smectic density and the
nematic director, respectively, to the quenched disorder of the
aerogel matrix, and are combined into $F_d=F_{d\rho}+F_{dn}$:
\begin{eqnarray}
F_d=\int d^d r \left [ \frac{1}{2} \delta t({\bf r})
(\rho-\rho_0)^2 + U({\bf r}) \rho +
\big({\bf g}({\bf r})\cdot{\bf n}\big)^2\right ] ,\label{Fd}
\end{eqnarray}
where both the quenched random $T_c$ ($\delta t({\bf r})$) and the
quenched random potential $U({\bf r})$ are proportional to the local
aerogel density $\rho_A({\bf r})$, and the quenched field ${\bf
g}({\bf r})$ is random and short-range correlated in direction, with
strength proportional to the local aerogel density. 

Combining the "random field" energy (first two terms of Eq.\ref{Fd})
with the relation between the smectic order parameter $\psi$ and the
density $\rho$, we obtain:
\begin{equation}
F_{d\rho}[\psi]=\int \! d^d{\bf r}
{1\over 2}\bigg[\delta t({\bf r})|\psi|^2 + V({\bf r})\psi +
V^*({\bf r})\psi^*\bigg]\;,
\label{Fdrho}
\end{equation}
where $V({\bf r}) \equiv U({\bf r})e^{iq_0z}$. Note that, despite the
long-ranged correlations of $U({\bf r})$, which arise due to the
fractal structure of the aerogel \cite{Clark}, $V({\bf r})$ has only
short-ranged correlations.  This is because correlations of $V$ near
$\bf q=\bf 0$ are related to those of $U$ near ${\bf q} = q_0
{\bf\hat{z}}$ and the aerogel itself has no particular spatial
structure at the wavevector of the smectic ordering, $q_0$.  Thus, the
correlations of $V({\bf r})$ are short-ranged, and hence we can
accurately capture the long distance physics by taking the real space
correlations to be zero-ranged, and write $\overline{V({\bf
r})V^*({\bf r'})} = \Delta_V \delta^d({\bf r}-{\bf r'})$ where
$\Delta_V=C_U(q_0 \hat{z})$ (overbar denotes quenched disorder
average). Expanding in small deviations from perfect nematic order
${\bf n}_0={\bf\hat{z}}$, writing ${\bf {\hat n}}({\bf r})={\bf {\hat
z}} + \delta {\bf n}({\bf r})$, to linear order in $\delta{\bf n}({\bf
r})$, $F_{dn}$ (the last term in Eq.\ref{Fd}) becomes:
\begin{equation}
F_{d n}\approx\int d^d r\; {\bf h}({\bf r})\cdot\delta{\bf n}\;,
\label{Fn1}
\end{equation}
where we have defined a quenched random tilt field 
${\bf h}({\bf r}) \equiv g_z({\bf r}){\bf g}({\bf r})$.

Since we expect ${\bf g}({\bf r})$ to have only short-ranged
correlations (with range of order the orientational persistence length
of the silica fibers), the above correlation function of the tilt
disorder should also be short-ranged.  Furthermore, it must be
isotropic.  These considerations lead to the following form for the
correlation function: $\overline{h_i({\bf r})h_j({\bf r'})} \equiv
\Delta_h\delta^d({\bf r}-{\bf r'})\;\delta_{ij}$, which is
short-ranged.

While it is tempting to directly analyse the model
$F=F_{dG}+F_{d\rho}+F_{dn}$ written in terms of the smectic order
parameter $\psi$, we will not do so here\cite{toPublish}. Our
motivation for this is that such a direct approach is {\em known} to
{\em incorrectly} predict the lower critical dimension in, e.g., the
random field Ising model (as well as, e.g. completely missing the
existence of the Kosterlitz-Thouless transition). Indeed, it proves to
also do so here. Instead, we proceed by {\em assuming} the existence
of smectic order and writing $\psi({\bf r})=|\psi_0|e^{i q_0 u({\bf
r})}$, with a uniform amplitude $|\psi_0|=const.$ and $u({\bf r})$ the
local displacement of the smectic layers from the perfect periodic
order.  Using this low-T ansatz in Eqs.\ref{Fd}--\ref{Fn1}, and
integrating over the nematic director fluctuations $\delta {\bf n}$,
results in the replacement $\delta{\bf n}\rightarrow
\bbox{\nabla}_\perp u$, everywhere in $F[u,\delta{\bf n}]$ (the Higg's
mechanism). This leads to the elastic free energy of the disordered
smectic-A phase, valid in the long wavelength limit, to quadratic
order in gradients of $u$, and provided dislocations are confined,
\begin{eqnarray}
F[{u}]=\int \! d^d{\bf r} 
\bigg[  {B\over 2}(\partial_z
{u})^2 + {K\over 2}(\nabla^2_\perp {u})^2
+ {\bf h}({\bf r})\cdot\bbox{\nabla_\perp} u\nonumber\\ 
-  |\psi_0|\left(V({\bf r})e^{i q_0 u({\bf r})}+
V^*({\bf r})e^{-i q_0 u({\bf r})}\right)
\bigg]\;.
\label{Fuii}
\end{eqnarray}

To compute self-averaging quantities, we employ the replica ``trick''.
After replicating and integrating over the disorder\cite{toPublish},
we obtain $\overline{Z^n}=\int[d u_\alpha]e^{-F[u_\alpha]}$, with
\begin{eqnarray}
&&F[u_\alpha]=\int d^d
r\bigg[\sum_{\alpha=1}^n\left({K\over2}(\nabla_\perp^2
u_\alpha)^2+{B\over2}(\partial_z u_\alpha)^2\right)\label{Fr}\\
&&+\sum_{\alpha,\beta=1}^n
\left({\Delta_h\over4}|\bbox{\nabla_\perp}(u_\alpha-u_\beta)|^2 -
\gamma\cos[q_0(u_\alpha-u_\beta)]\right)\bigg]\;,\nonumber
\end{eqnarray}
where $\gamma\equiv|\psi_0|^2\Delta_v/2$. We have studied this model using
the standard momentum shell renormalization group (RG) transformation,
generalized to allow for anisotropic scaling:
$r_\perp=r_\perp'e^{\ell}$, $z=z'e^{\omega\ell}$\cite{toPublish}; the
results are the RG flow equations in 3d:
\begin{eqnarray}
\frac{d\gamma(\ell)}{d\ell}&=&(2+\omega-\eta)\gamma - A_1\gamma^2\;,
\label{g}\\
\frac{d K(\ell)}{d\ell}&=&(\omega-2)K\;,
\label{K}\\
\frac{d B(\ell)}{d\ell}&=&(2-\omega)B\;,
\label{B}\\
\frac{d\Delta_h(\ell)}{d\ell}&=&\omega\Delta_h + A_2\gamma^2\;,\label{Deltah}
\end{eqnarray}
where for simplicity we set the UV cutoff $\Lambda=1$, and $\eta =
q_0^2/(4\pi\sqrt{K B})$, $A_1 = 2\eta/(3\pi K)$, $A_2 = q_0^2\eta/(4 K)$.
The statistical symmetry under global rotation requires that the
disorder generated replica off-diagonal terms to be invariant under
$u_\alpha({\bf r})\rightarrow u_\alpha({\bf r})+ \bbox{\theta}\cdot
{\bf r}_\perp$. In Eq.\ref{Fr} the nonlinearities only depend on the
difference between different replica fields and therefore do not
depend on the ``center of mass'' field $\sum_{\alpha=1}^n u_\alpha$,
which is therefore a noninteracting field.  This implies that $K$ and
$B$ are {\em not} renormalized by disorder,\cite{CO,TD} i.e. their
flow equations are {\em exact}, ignoring (for now) the effects of both
anharmonic elastic terms and topological defects loops in $u$.  Note
that $\eta$, which is simply the Caille exponent for the algebraic
decay of smectic correlations in the pure smectic, is
unrenormalized. The recursion relation for the proper dimensionless
coupling constant $\tilde{\gamma}\equiv 2\gamma/(3\pi K)$ can be easily
obtained by combining Eqs.\ref{g},\ref{K}
\begin{eqnarray}
\frac{d\tilde{\gamma}(\ell)}{d\ell}&=&(4-\eta)\tilde{\gamma} -
\eta\;\tilde{\gamma}^2\;,
\label{g2}
\end{eqnarray}
and, as required, is independent of the arbitrary anisotropy rescaling
exponent $\omega$. From Eq.\ref{g2}, we then find that for $\eta<4$
(large elastic moduli), away from the NA transition $K
B>q_0^4/256\pi^2$, the smectic fixed line is unstable to
disorder. However, this instability to disorder is stabilized by the
nonlinear terms in $\tilde{\gamma}$, which terminate the flow at a new
finite disorder fixed line, $\tilde{\gamma}^*=(4-\eta)/\eta$.  This
new fixed line then controls a glassy smectic-A phase, analogous to
the super-rough phase of crystal surface on a random
substrate\cite{CO,TD} and the vortex glass phase of flux-line vortices
in type II superconductors\cite{CO,VG}.

The flow Eq.\ref{g2} also implies that the random-field disorder is
irrelevant for $\eta>4$. Since the bulk modulus $B$ vanishes while $K$
remains finite through $T_{NA}$, $\eta$ diverges as $T \rightarrow
T^-_{NA}$, and hence near the bulk NA transition T$_{NA}$, we are {\em
guaranteed} to have a range of $T$ over which the random-field
disorder is irrelevant. However, as we will see below, because tilt
disorder ($\Delta_h$) is a strongly relevant perturbation, the 3d
quasi-long-range smectic order for $\eta>4$ will be converted into
short-range correlations, even when the random field disorder given by
$\gamma$ is irrelevant.  It is essential to stress that the RG flow
described above (i.e. relevance for $\eta<4$ and irrelevance for
$\eta>4$ of the random-field disorder) survives even despite the
strong relevance and runaway of the random tilt coupling $\Delta_h$.

As can be seen from the recursion relations Eq.\ref{Deltah}, even if
the bare $\Delta_h=0$, tilt disorder is generated by the random-field
disorder $\gamma$ upon renormalization.  In contrast to the 2d
random-field XY-model, where the generated $\Delta_h$ disorder is only
marginally relevant and only weakly affects the QLRO order found for
$\Delta_h=0$ (replacing $\log$ phase correlations by
$\log^2$)\cite{CO,TD}, for the 3d smectic-A phase $\Delta_h$ tilt
disorder is strongly relevant.  The effect of the tilt disorder is
controlled by the dimensionless coupling
$g\equiv{\Delta_h}/({B\lambda^3})$, where $\lambda\equiv(K/B)^{1/2}$.
From the recursion relations Eqs.\ref{g}-\ref{Deltah} we find: ${d
g}/{d\ell}=2g+(9\pi^3\eta^2/4){\tilde\gamma}^2$.  For $\eta > 4$,
${\tilde\gamma}(\ell)\rightarrow 0$ (as we have seen), and so $d
g/d\ell=2g$, which is trivially solved to give $g(\ell)=g(0)
e^{2\ell}$, where $g(0)$ is a constant.  Thus, the tilt disorder {\em
is} strongly relevant. We expect this on physical grounds since random
tilt disorder explicitely breaks the rotational invariance of the
smectic-A phase.

For $\eta < 4$, ${\tilde\gamma}\rightarrow {\tilde\gamma}^* > 0$. Now,
in the 2d random-field XY model, the existence of a non-zero
${\tilde\gamma}^*$ in the low-T phase implied completely different
behavior for the tilt coupling $g(\ell)$ than in the high-T phase.  In
{\em our} problem, however, solving Eq.\ref{g} gives
$g(\ell)=(g(0)+9\pi^3\eta^2{\tilde\gamma}_*^2/8)e^{2\ell}-
9\pi^3\eta^2{\tilde\gamma}_*^2/8$, which asymptotically runs away to
$+\infty$ as $\sim e^{2\ell}$ in {\em both} phases, {\em in exactly the
same way}\ !  Non-universal constants (like $g_0$) change but the
scaling ($e^{2\ell}$) does not.  This implies that {\em equal time}
correlation functions scale in exactly the same way in both the
``glassy '' $(\eta < 4)$ and non-glassy $(n > 4)$ phase.  We will
therefore calculate them in the non-glassy ($\eta>4$) phase, where we
can set $\gamma=0$; our results, however, will apply to {\em both}
phases. The only difference between these two phases is that the {\em
dynamics} are slower in the glassy phase.

For $\gamma=0$ we can calculate anything. The quantity
of interest $C({\bf r}_\perp,z)=\overline{\langle\left(u({\bf
r_\perp},z)-u({\bf 0},0)\right)^2\rangle}$ is
\begin{equation}
C({\bf r}_\perp,z)=2 \int\frac{d^2q_\perp dq_z}{(2\pi)^3}(1-e^{i{\bf
q}\cdot{\bf r}})\overline{\frac{\langle u({\bf q})u({\bf
q}')\rangle}{\delta^d({\bf q}+{\bf q}')}}\;,
\label{C1}
\end{equation}
where the quenched and thermal averaged $\overline{\langle u({\bf
q})u({\bf q}')\rangle}= \langle u_\alpha({\bf q})u_\alpha({\bf
q}')\rangle =\delta^d({\bf q}+{\bf q}')G_{\alpha\alpha}({\bf q})$ is
expressed in terms the replicated correlation function, where no sum
on $\alpha$ is implied, and the replica propagator
$G_{\alpha\beta}=\delta_{\alpha\beta}/\Gamma_q+\Delta_h q_\perp^2
/\Gamma_q^2$, with $\Gamma_q\equiv K q_\perp^4+B q_z^2$, can be read
off from Eq.\ref{Fr}.  This together with Eq.\ref{C1} gives, in the
${\bf r}\rightarrow\infty$ limit,
\begin{eqnarray}
C&=&2\Delta_h\int{d^2{q_\perp}d
q_z\over(2\pi)^3}{q_\perp^2\left(1-e^{i{\bf q}\cdot{\bf r}}\right)
\over\left(K q_\perp^4 + B q_z^2\right)^2}\;,\label{CD}\\
&=&\frac{\Delta_h}{32\pi B\lambda^3}
\left\{4\lambda|z| e^{-r_\perp^2/4\lambda|z|}\right.\nonumber\\
&+&\left.r_\perp^2\left[Ei\left({-4\lambda L_z\over L_\perp^2}\right)+
Ei\left({-r_\perp^2\over4\lambda|z|}\right)
+2\log\left({L_\perp\over r_\perp}\right)\right]\right\}\;,\nonumber
\end{eqnarray}
where $Ei(x)$ is the exponential integral function, we have considered
a finite system whose shape is a rectangular parallelepiped of linear
dimensions $L_\perp\times L_\perp\times L_z$, $L_z$ being the length
of the system along the ordering ($z$) direction, and dropped the
subdominant thermal (finite $T$) contribution to $C({\bf r}_\perp,z)$.
In the usual $\lambda L_z << L_\perp^2$ limit, the asymptotic
behaviors of $C({\bf r}_\perp,z)$ are
\begin{eqnarray}
C\approx\left\{\begin{array}{lr}
\frac{\Delta_h}{32\pi B^2\lambda^3}
\left[4 \lambda|z|+r_\perp^2\log|\frac{L_z}{z}|\right],
&\lambda|z|>>r_\perp^2,\\
&\lambda L_z<<L_\perp^2\label{limit1}\\
\frac{\Delta_h}{16\pi B^2\lambda^3}\,
r_\perp^2\log\left(\frac{2\sqrt{\lambda L_z}}{r_\perp}\right), 
&\lambda |z|<<r_\perp^2,\\
&\lambda L_z<<L_\perp^2\label{limit2}
\end{array}\right.
\end{eqnarray}

An unusual feature of this result is that even the {\em relative}
displacement of two points with {\em finite} separations $(r_\perp,
z)$ diverge as the system sizes $(L_\perp, L_z)$ go to infinity.  This
is because the mean squared real space {\em orientational}
fluctuations $\overline{\langle|\delta{\bf n}({\bf
r})|^2\rangle}=\overline{\langle|\bbox{\nabla}_\perp u({\bf
r})|^2\rangle}$ also diverge as $L_{\perp,z}\rightarrow\infty$:
\begin{eqnarray}
\langle |\delta{\bf n}({\bf r})|^2\rangle 
&=& 2\Delta_h\int{d^2{q_\perp}d q_z\over(2\pi)^3}
{q_\perp^4\over\left(K q_\perp^4 + B q_z^2\right)^2}\;,\nonumber\\
&=&{\Delta_h\over4\pi B^2\lambda^3}
\log\left({\mbox{min}}[\sqrt{\lambda L_z},L_\perp]\right)\;.
\label{orientation}
\end{eqnarray}

Defining the {\em translational} correlation lengths $\xi_\perp$ and
$\xi_z$, as the distances $r_\perp$ and $z$ at which $C({\bf
r}_\perp,z)$ is of order $a^2$, where $a$ is a lattice constant, gives:
$\xi_z=a^2 8\pi K B/\Delta_h$, and
\begin{equation}
\xi_\perp=4a\left({\pi B^2\lambda^3\over\Delta_h
\log(2\sqrt{\lambda L_z}/\xi_\perp)}\right)^{1/2}
\label{xiDeltap}
\end{equation}

Furthermore, because the liquid crystal in aerogel lacks long-ranged
{\it orientational} order, as well, to obtain the X-ray scattering
from aerogel, one must powder average; the broad {\em ring} of X-ray
scattering that results has width
$\kappa_{powder}\cong(\xi_z)^{-1}={\Delta_h}/(8\pi B K a^2)$.

The {\em orientational} correlation lengths $\xi_{\perp,z}^{o}$, can
similarly be defined as the values of $L_{\perp,z}$ beyond which the
mean squared orientational fluctuations $\overline{\langle|\delta{\bf
n}|^2\rangle}$ of Eq.{} get to $O(1)$.  This gives: $\xi^o_\perp=a
e^{4\pi B^2\lambda^3/\Delta_h}= a e^{\xi_z\lambda/2a^2}$,
$\xi^o_z=(a^2/\lambda)e^{2\pi B^2\lambda^3/\Delta_h}
=(a^2/\lambda)e^{\xi_z\lambda/4a^2}$.  Thus, orientational
order persists out to {\em much} larger distances than translational
order, in the limit of weak disorder where {\em all} the correlation
lengths get large.

All of the above results apply subject to our two initial assumptions,
that: (1) dislocations were {\em not} generated by the disorder, and
(2) anharmonic terms in the elastic free energy could be neglected.

We will show now that, if we continue to assume (2) (whose validity we
will investigate in a future publication\cite{toPublish}), assumption
(1) is wrong: in the {\em harmonic} elastic approximation,
dislocations {\em are} created even by arbitrarily weak disorder.
However, they are felt only on length scales longer than
$\xi^o_{\perp,z}$, and hence much longer than the translational
correlation lengths.  Thus, our above calculations of these lengths
remain valid.

We can include dislocations in the ``tilt only'' model,
i.e. Eq.\ref{Fuii} with $V({\bf r})=0$.  As discussed earlier, this
theory correctly reproduces all of the static correlation functions in
{\em both} the glassy and the non-glassy regimes.

The dislocations are characterized by an integer-valued 3d vector
field ${\bf m}({\bf r})$ defined on the sites ${\bf r}$ of a lattice
connected to the displacement field $u$ via: $\bbox{\nabla}\times{\bf
v}={\bf m}$, with ${\bf v}\equiv\bbox{\nabla}u$, and a dislocation
line continuity constraint $\bbox{\nabla}\cdot{\bf m}({\bf r})=0$.
Standard manipulations\cite{toPublish,Toner}, lead to a Coulomb gas
theory of these dislocation loop defects:
\begin{eqnarray}
H_d={1\over2}\int_{\bf q}\left[{K q_\perp^2\over\Gamma_q} P_{i j}^\perp
m_i({\bf q}) m_j(-{\bf q}) + {\bf m}({\bf q})\cdot{\bf a}(-{\bf q})\right],
\label{Hd}
\end{eqnarray}
where the inverse of the pure smectic propagator $\Gamma_q\equiv q_z^2
+ \lambda^2 q_\perp^4$, $P_{i j}^\perp({\bf q})=\delta_{i j}^\perp -
q_i^\perp q_j^\perp/q_\perp^2$, and ${\bf a}({\bf q})$ is a Fourier
transform of the quenched field related to the original random phase
field ${\bf h}({\bf q})$ via $-i{\bf a}= {\bf q}\times {\bf h}/q^2 -
\left (\hat{\bf z} \times {\bf q} \right ){\bf q}\cdot{\bf h}\, q_z
\left(1-\lambda^2q^2_{\perp} \right)/({\Gamma_q q^2})$.

The partition function for this model is then $Z[\{{\bf
h}\}]=\sum_{\{{\bf m}({\bf r})\}}^\prime e^{-S[\{{\bf m}\}]}$, where
$S\equiv H_d/T + E_c/T\sum_{\bf r}|{\bf m}({\bf r})|^2$, and the sum
is over all integer-valued configurations of ${\bf m}$, satisfying
the dislocation line continuity constraint $\bbox{\nabla}\cdot{\bf
m}=0$, $H_d$ is given by Eq.\ref{Hd}, and we have added a core energy
term $E_c\sum_{\bf r}|{\bf m}({\bf r})|^2$.  To proceed, we enforce
the constraint $\bbox{\nabla}\cdot{\bf m}=0$ by introducing a new
auxiliary field $\phi({\bf r})$, and introduce a dummy gauge field
$\bf A$ to mediate the long-ranged interaction between defects loops
in the Hamiltonian Eq.\ref{Hd} obtaining
\begin{eqnarray}
Z=\prod_{\bf r}\int d\phi({\bf r})d{\bf A}({\bf r})
\sum_{\{{\bf m}({\bf r})\}} e^{-S[\{{\bf m}\},\phi,{\bf A}]}
\delta(\bbox{\nabla}\cdot{\bf A})\delta(A_z)\;,\nonumber\\
\nonumber\label{Z3}
\end{eqnarray}
\begin{eqnarray}
S&=&\beta\sum_{\bf r}\bigg[{\bf m}
({\bf r})\cdot\big(-i\bbox{\nabla}\phi({\bf r})+i{\bf A}({\bf
r})+{\bf a}({\bf r})\big) + E_c|{\bf m}|^2\bigg]\nonumber\\
&+&{1\over2}\sum_{\bf q}{\Gamma_q\over K q_\perp^2}|{\bf A}|^2\;,
\label{S2}
\end{eqnarray}
Performing the summation of the dislocation loop degrees of freedom,
replacing the resulting Villain potential by a cosine, generalizing to
a ``soft spin'' model described by a ``disorder'' parameter
$\psi=|\psi|e^{i\phi}$, we finally obtain\cite{toPublish} a complex
action $S$
\begin{eqnarray}
S&=&\sum_{\bf r}\bigg[\big(\bbox{\nabla}+i{\bf A}+{\bf a}\big)\psi^*
\big(\bbox{\nabla}-i{\bf A}-{\bf a}\big)\psi
+t|\psi|^2+u |\psi|^4\bigg]\nonumber\\ 
&+&\sum_{\bf q}{\Gamma_q\over 2 K q_\perp^2} |{\bf A}({\bf q})|^2\;,
\label{S4}
\end{eqnarray}

A complete discussion of the behavior of the above
model\cite{toPublish} is outside the scope of this Letter; here we are
only interested in the question of whether the dislocations loops will
or will not unbind at any, even infinitesimal amount of disorder.
Using replicas and computing the disorder-averaged free energy, we
find that the lowest order contribution to the renormalized {\em dual}
temperature $t_R$ comes from the average of the ``diamagnetic'' term
$\delta S=\sum_{\bf r}\big(\langle|{\bf A}|^2\rangle - \overline{|{\bf
a}|^2}\big)|\psi|^2$, which gives
\begin{equation}
t_R=t_0 + (d-2)\int{d^d q\over(2\pi)^d}\left[{K
q_\perp^2\over\Gamma_q} - {\Delta_h q_z^2 q_\perp^2\over q^2
\Gamma_q^2}\right],
\label{tR}
\end{equation}
where we have used the connection between ${\bf a}({\bf r})$ and the quenched
tilt disorder ${\bf h}({\bf r})$, averaged over ${\bf h}({\bf r})$ using its
distribution function, and generalized to $d$-dimensions.

The second, disorder, term in this integral dominates the first as
${\bf q}\rightarrow 0$. Indeed, this integral diverges in the
infra-red for $d\leq 3$
\begin{eqnarray}
\int d^d q {q_z^2 q_\perp^2\over q^2(q_z^2 + \lambda^2
q_\perp^4)^2}
\propto \int {d^{d-1}q_\perp\over q_\perp^2}\;,
\label{diverge}
\end{eqnarray}
where we used the fact that the dominant regime of the integral is
$q_z\sim\lambda q_\perp^2$. This divergence implies that $t_R$ is
driven to $-\infty$ (note the minus sign) by the disorder in
$d=3$. Indeed, we find in 3d $t_R=t_0 -
\frac{\Delta_h}{\lambda}\log(L/a)\times O(1)$, where $L$ is an IR
cutoff (e.g., the lateral extent of the smectic layers) and $a$ is the
UV cutoff (e.g., $\sim10$\AA, the size of the liquid crystal molecules).

This implies that the {\it dual} (dis-)order parameter is always in
its ordered phase, which, in turn, implies that the dislocation loops
of the original smectic model are always {\it unbound}, thereby
destroying the smectic order, even at $T=0$, for any infinitesimal
amount of disorder.

These results of the {\em harmonic} theory imply that for 3d
disordered smectics the dislocations are unbound, even at $T=0$, for
arbitrarily weak disorder. This means that there is no
thermodynamically {\em sharp continuous} NA transition, and the low
temperature phase must have a {\em finite} smectic translational
correlation length (even at $T=0$), once the liquid crystal is put in
aerogel. A {\em first} order transition between nematics with short
and long (but finite) smectic correlation lengths, or a smeared
analytic crossover, are of course always possible and have been
observed in experiments of Refs.\cite{HG} and
\cite{Clark}, respectively. We have also shown\cite{toPublish} that
once the dislocations are unbound, the static {\it director}
fluctuations are precisely those of a {\it nematic} in a random tilt
field. This implies that the system can be thought of as a nematic in
such a random field, at all temperatures and all disorder
strengths. This, in turn, can be shown\cite{toPublish} to imply that
long-ranged orientational (nematic) order is destroyed as well, again
in agreement with experiments\cite{Clark}.

Finally, preliminary investigation of anharmonic elastic effects suggests
that they {\em may} prevent dislocation unbinding and stabilize
orientational order.\cite{toPublish} Furthermore, our results {\em do
not} imply that there is no thermodynamically sharp phase transition in
this system analogous to the {\it nematic to isotropic} phase transition
in the pure system. Indeed, preliminary investigation suggests that a
kind of "nematic Bragg glass" phase may exist in these systems. This
possibility will be further discussed in a future
publication\cite{toPublish}.

L.R. and J.T. thank N. Clark for discussions and acknowledge financial
support by the NSF through Grants DMR-9625111 and DMR-9634596,
respectively.

\end{multicols}

\vspace{-.2in}
\begin{references}
\vspace{-.7in}
\bibitem{Clark} N.A. Clark, et al., Phys. Rev. Lett. {\bf 71}, 3505
(1993). T. Bellini, et al., ibid., {\bf 74}, 2740 (1995).
\bibitem{HG} H. Haga, C. W. Garland, Liq. Cryst. {\bf 22}, 275 (1997).
\bibitem{toPublish} L. Radzihovsky and J. Toner, Phys. Rev. Lett. {\bf 78},
4414 (1997), and unpublished.
\bibitem{BraggGlass} T. Giamarchi, et al., Phys. Rev. Lett. {\bf 72},
1530 (1994).
\bibitem{deGennes} P. G. de Gennes, Solid State Comm. {\bf 10}, 753
(1972).
\bibitem{CO} J. Cardy and S. Ostlund, Phys. Rev. B {\bf 25}, 6899 (1982).
\bibitem{TD} J. Toner, D. DiVincenzo, Phys. Rev. B {\bf 41}, 632
(1990).
\bibitem{VG} M. P. A. Fisher, Phys. Rev. Lett. {\bf 62}, 1415 (1989);
\bibitem{Toner} J. Toner, Phys.Rev.B {\bf R26}, 462 (1982).
\end{references}
\end{document}